\documentclass[letterpaper, preprint, paper,11pt]{AAS}	

\usepackage{bm}
\usepackage{amsmath}
\usepackage{subfigure}
\usepackage{subcaption} 
\usepackage{svg}
\usepackage[colorlinks=true, pdfstartview=FitV, linkcolor=black, citecolor= black, urlcolor= black]{hyperref}
\usepackage{overcite}
\usepackage{footnpag}			      	
\usepackage{hyperref}
\usepackage{footmisc}
\usepackage{svg}
\usepackage{tabularx}
\usepackage{threeparttable}
\usepackage{longtable}
\usepackage{graphicx}
\usepackage{comment}

\PaperNumber{25-744}

\begin{document}

\newcommand{\affilA}{\footnotemark[1]} 
\newcommand{\affilB}{\footnotemark[2]} 
\title{STELLAR OCCULTATIONS IN SUPPORT OF THE LUMIO ORBIT DETERMINATION }
\author{Davide Banzi\affilA,
Riccardo Lasagni Manghi\affilA\ \affilB,
Marco Zannoni\affilA\ \affilB
}

\maketitle{} 		
\footnotetext[1]{Department of Industrial Engineering (DIN), University of Bologna, via Fontanelle 40, 47121, Forlì, Italy. davide.banzi4@unibo.it} 

\footnotetext[2]{Centro Interdipartimentale di Ricerca Industriale Aerospaziale (CIRI AERO) Alma Mater Studiorum – Università di Bologna, Via Baldassarre Carnaccini 12, 47121 Forlì, Italy.} 

\begin{abstract}
This work investigates the use of stellar occultation measurements to enhance the orbit determination performance of the Lunar Meteoroid Impact Observer (LUMIO) mission, operating from a quasi-Halo orbit around the Earth-Moon L2 point. During science phases, when radiometric tracking is sparse and low illumination limits conventional optical navigation methods, occultation events, defined as precise timings of stellar appearances/disappearances behind the Moon’s limb, offer a suitable alternative.
A simulation tool based on JPL’s MONTE library was developed to identify valid occultation events, applying geometric and illumination constraints to exclude non-observable cases. These events were integrated into a batch least-squares orbit determination filter alongside conventional radiometric data. The covariance analysis shows that occultation observables reduce the transverse and normal position uncertainties of LUMIO by up to a factor of two, especially during tracking gaps or occultation-rich arcs. This uncertainty reduction is expected to facilitate station-keeping operations and constrain the surface localization of Lunar Impact Flashes (LIFs), enhancing the mission's scientific return.
Sensitivity analyses confirm that the orbit determination performance is primarily driven by the timing accuracy of occultation events, with limited dependence on lunar shape uncertainty below 100 m. These findings confirm the potential of occultation-based navigation to enhance spacecraft autonomy and robustness in low-visibility environments, making it a valuable complement to radiometric techniques for future lunar and deep-space missions.

\end{abstract}

\section{Introduction}

The Orbit Determination (OD) process is used for real-time navigation of deep space probes during mission operations and allows for the reconstruction and prediction of the probe's trajectory. Furthermore, a precise orbit determination is essential to associate in situ and remote scientific observations of the mission target with an absolute position in space.

While missions that occur near the Earth or in a cislunar environment often rely on global navigation satellite systems for positioning, deep space missions mainly rely on ground-based radiometric measurements, namely Doppler, range, and Delta-DOR, to estimate the spacecraft’s state. Since the 1950s, deep space probes have also relied on measurements derived from optical images of planetary and small-body targets.

Optical measurements include, among others, geometric centers of convex shapes fitted to the target limb, sample and line coordinates of known features on the target's surface \cite{christian2009review, pellacani2019hera, de2015optical, rush2023optical}, or angular separation between target landmarks and background stars \cite{van1982voyager, jorris1967experiment}. Optical methods are particularly useful when a precise relative position with respect to the target is needed (i.e., during flybys, proximity operations, and landing) or when a high level of autonomy from ground operations is needed.
However, these methods may fail in providing an accurate spacecraft state estimation for high values of Sun phase angle, when significant portions of the target's surface are in shadow \cite{franzese2019autonomous}.

An alternative technique that may overcome this limitation exploits the time of star occultations behind the target's limb to constrain the spacecraft's position. This method has been previously studied for the Gemini X mission, where the recorded times of occultation allowed to reach spacecraft position uncertainties down to 1.8 km \cite{keenan1962star, jorris1967experiment}.
More recently, Psiaki and Hinks \cite{psiaki2007autonomous} proposed an extended Kalman filter, based on measurements of star occultation/rising times, for autonomous orbit determination in lunar orbits. Ground-truth simulations, assuming a measurement noise of 1 ms and uncertainty in the Moon's shape of 100 m, achieved steady-state peak position uncertainty of 70 m for each axis. 
Similarly, Landgraf et al. \cite{landgraf2006optical} proposed using times of star occultations for human lunar exploration missions. Specifically, they used orbital mechanics and geometry to determine the true anomaly at which an occultation event should occur. Sensitivity analyses for a satellite approaching the Moon hyperbolically yielded precisions of the peri-selenium altitude better than 1 km, assuming a 24-hour data cutoff and measurement errors of less than 10 s. 

This work provides a feasibility analysis for using stellar occultation measurements in support of the Lunar Meteoroid Impact Observer (LUMIO) mission. We want to assess if complementing Earth-based radiometric data with stellar occultation measurements improves navigation accuracy during science observation windows, when traditional optical measurements are less effective. Previous studies modeled the occultation events using specific geometrical conditions as proxies. For example, Psiaki and Hinks defined occultations by measuring the minimum Lunar altitude along the line of sight vector from the spacecraft to the star. This choice was made due to limitations imposed by the extended Kalman filter, where a measurement model based on time observables (representing the free variable) is challenging to implement. Similarly, Landgraf modeled the eclipse events as zeros of an analytic function of the true anomaly. 
By using a direct time measurement of the occultation event, this work has a more intuitive formulation, which allows for quantifying different noise sources that can affect the occultation measurements.
Furthermore, we overcome limitations of previous works by incorporating dynamical and observational model uncertainties directly within the batch least-squares filter. These include uncertainties in the Moon's shape model, star coordinates, gravitational field, and onboard clock.
Finally, we explore a new mission scenario of a satellite orbiting the L2 Lagrangian point of the Earth-Moon system in a quasi-Halo orbit.

The first section of this paper describes the most relevant features of the LUMIO mission. The second section introduces the simulation procedure, focusing on the measurement model and defining the dynamical model and the filter setup used for the analysis. The third section shows the main results obtained from a covariance analysis. Initially, formal state covariances are presented and compared for two alternative scenarios, one using only radiometric measurements within the OD setup, and one adding stellar occultation observables. Furthermore, sensitivity analyses assess the robustness of the setup by showing uncertainty variations as a function of the main performance drivers, namely measurement noise and \textit{a priori} knowledge of the Moon's shape. Finally, the last section draws the conclusions and outlines future investigation topics.

\section{Mission Scenario}

LUMIO is a lunar exploration mission funded by the European Space Agency and scheduled to launch in 2027 \cite{cervone2022lumio}. The probe will operate for one year in a quasi-Halo orbit around the Earth-Moon Lagrangian point L2 in a 2:1 resonance with the synodic period. From this vantage point, the probe can continuously observe the lunar far side with its primary payload camera, LUMIO-Cam, looking for Lunar Impact Flashes (LIF) caused by meteoroid impacts on the surface. LUMIO's primary goal is to improve the understanding of the lunar meteoric environment and situational awareness for future lunar missions \cite{topputo2025lumio}, complementing ground-based LIF observations of the Moon's near side. Table \ref{tab:my_label1} summarizes the LUMIO mission phases.

\begin{table}[h]
    \centering
    \caption{Phases of the LUMIO mission.}
    \label{tab:my_label1}
    \begin{tabularx}{\textwidth}{|l|X|}
        \hline \textbf{Phase name} & \textbf{Description} \\
        \hline
        \hline 
        \textbf{Launch and early operations phase} & LUMIO is inactive in the deployer until the trans-lunar injection.\\
        \hline 
        \textbf{Commissioning phase} & LUMIO is deployed 0.5 days after the trans-lunar injection and is inserted into a Weak Stability Boundary transfer orbit. Commissioning operations are performed.\\
        \hline
        \textbf{Transfer phase} & During the orbit transfer, deep space maneuvers are performed. The 
        phase ends with a Halo injection maneuver. \\
        \hline
        \textbf{Operative phase} & Science and Navigation and Engineering 
        cycles are performed alternately for about one year. \\
        \hline
        \textbf{End-Of-Life phase} & LUMIO is sent in a re-entry orbit on the Moon. \\
        \hline 
    \end{tabularx}
\end{table}

\begin{figure}[h]
    \centering
    \includegraphics[width=0.5\textwidth]{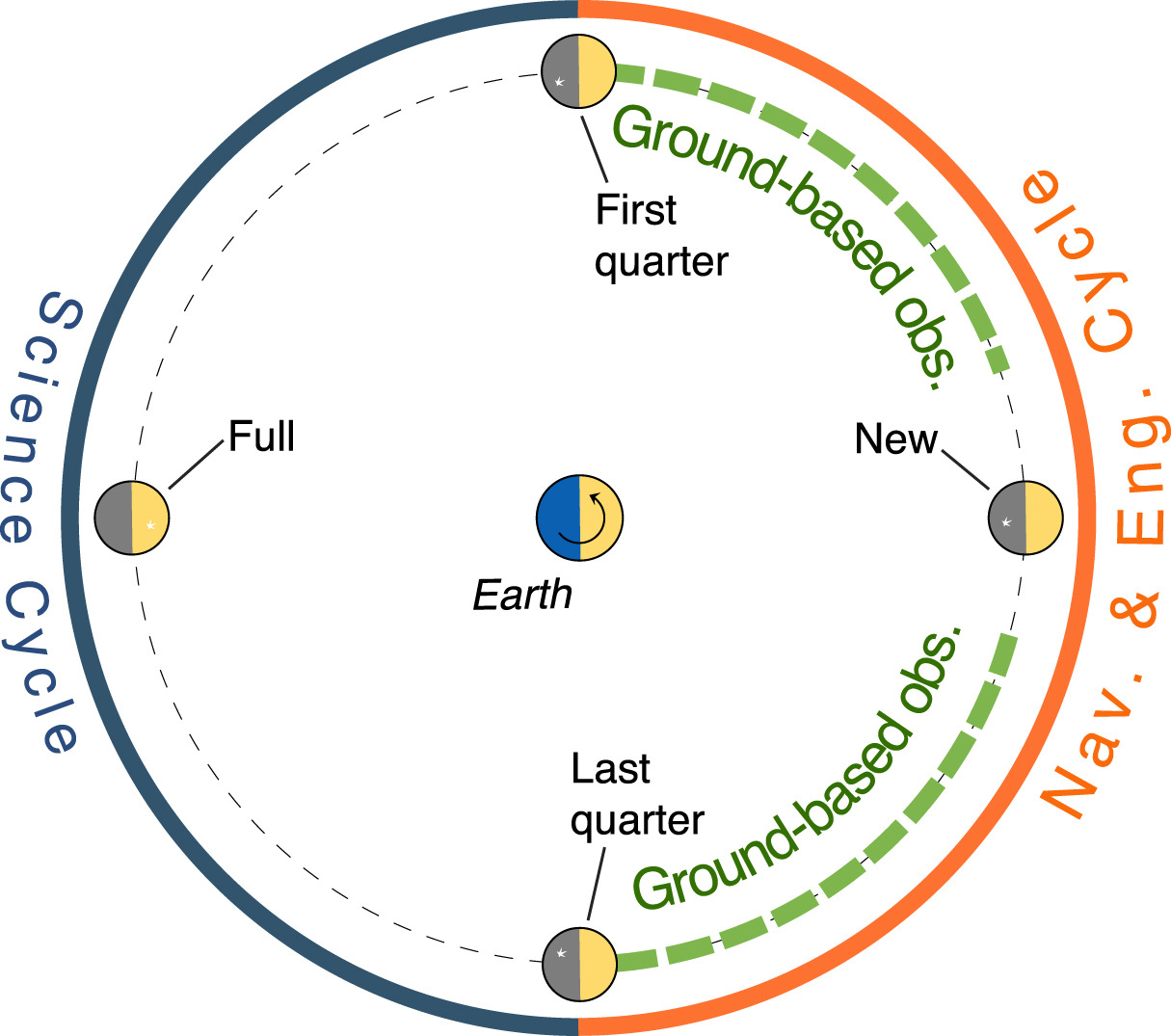}
    \caption{Cycles of LUMIO operative phase \cite{topputo2025lumio}.}
    \label{fig:mesh1}
\end{figure}

The operative phase of the mission will last 1 year and will be split into two cycles defined by the Sun's phase angle, as depicted in Figure \ref{fig:mesh1}. Scientific observations will occur during the Science Cycle, when the Sun phase angle is above 90 degrees and the Moon's far side is mostly in shadow, allowing for the observation of impact flashes. Conversely, Navigation and Engineering cycles correspond to periods during which the Sun phase angle is below 90 degrees. During these periods, LIF observations are complicated by the light reflected by the Moon's surface, and the main activities are related to station-keeping maneuvers and telecommunication operations.

\section{Simulation Procedure} 
This analysis is meant to support the LUMIO mission during its scientific operations, providing a preliminary estimation of the navigation performance when adding occultation measurement to standard radiometric observables.
To reach this goal, a numerical simulation of the OD process was performed. OD is an iterative process that begins with an \textit{a priori} estimate of the spacecraft trajectory. Precise measurement and dynamical models are implemented to compute the expected values of observables (computed observables) based on the spacecraft's nominal state. These computed observables are then compared with actual observed data (observed observables) to compute the residuals \cite{thornton2003radiometric}. A weighted least-squares filter is then applied to determine the values of dynamical and observational parameters that minimize the residuals, reflecting errors in the estimation.

In particular, we opted for a multi-arc approach, which involves segmenting the spacecraft's trajectory into distinct, non-overlapping intervals called arcs. Measurements from these separate arcs are collectively analyzed using a weighted least-squares batch filter, resulting in a unified set of estimated parameters \cite{godard2017multi}. Due to the particularity of the multi-arc approach in this work, a LUMIO science cycle coincides with one filter arc. 

Since LUMIO is still under development, synthetic occultation observables are generated using an occultation simulator and treated in the estimation filter like observed observables. The measurement model uses the same simulator to retrieve the computed occultation observables. The Mission Analysis, Operations, and Navigation Toolkit Environment (MONTE) Python library was used to implement the OD setup \cite{evans2018monte}. 

\subsection{Measurement Model}
\begin{figure}[h]
    \centering
    \includegraphics[width=1\textwidth]{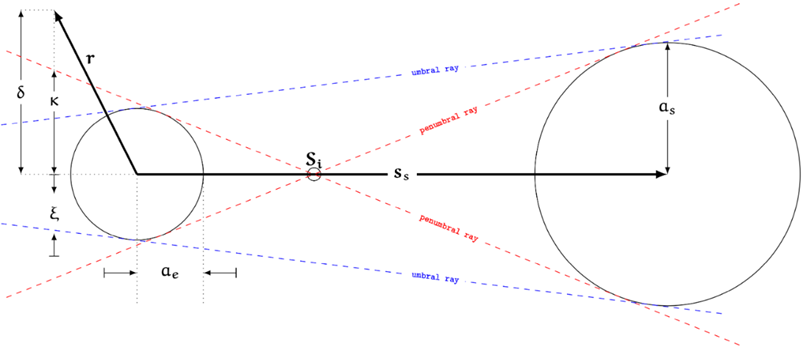}
    \caption{Graphical representation of the occultation geometry and its related parameters \cite{betts2015optimal}. In this analysis, the Moon is the occulting body (pedix e), stars are the occulted bodies (pedix s), and LUMIO is the observer }
    \label{fig:mesh2}
\end{figure}

The star occultations simulator computes the precise time of ingress and egress of a given star behind the Lunar limb, for which a spherical shape was assumed. Although not sufficiently accurate for the actual OD during operations, the use of a simplified shape does not affect the reliability of the analysis. The sensitivity properties of this model, which is used both for simulating the observables and for the estimation, are similar to those of a high-fidelity model. However, for the analysis of real data, accurate topographic maps of the Moon's surface and attitude cosine matrices of its body-fixed frame will be needed.
The event search relies on the \textit{OccultationEvent} function of the MONTE library, whose detailed mathematical formulation is described by Betts\cite{betts2015optimal}. Figure \ref{fig:mesh2} gives a graphical representation of the geometry of the problem and the relevant variables. In this work, only umbral rays are considered and assumed to be parallel because of the huge distances between the occulted stars and the Moon.
An occultation event occurs when the spacecraft's distance from the center of the umbral cone $\delta$ is equal to the distance between the umbral cone and the umbral terminator $\xi$. It is also possible to distinguish whether the occultation is in ingress or in egress. If the partial derivative of the equation $E(t)=\|\delta\|-\xi=0$ is positive, an ingress is observed. If it is negative, an egress is recorded instead.

For the baseline scenario, we considered only stars having a visual magnitude lower than 6, which is consistent with the current sensor specifications of the LUMIO-Cam \cite{cervone2022lumio} and its assumed sampling rate of 60 ms from the preliminary concept of operations \cite{merisio2020predicting}.
The stellar catalog used in the simulator is UCACT-PI, which merges the UCAC2 \cite{zacharias2004second} and Tycho-2 \cite{hog2000tycho} catalogs while using the parallax information from Hipparcos \cite{perryman1997hipparcos}. This catalog, which was used for both measurement simulation and estimation within the filter, has a good star coverage in the selected magnitude range. However, future studies may explore alternative catalogs by performing a trade-off between stars position accuracy and spatial coverage, in preparation for the actual data analysis.

To account for the influence of stray light from the Earth and the Sun, we discarded all events occurring when the two bodies had an angular separation of less than 17.5 degrees from the LUMIO-Cam boresight.

\begin{figure}[h]
    \centering
    \includegraphics[width=1\textwidth]{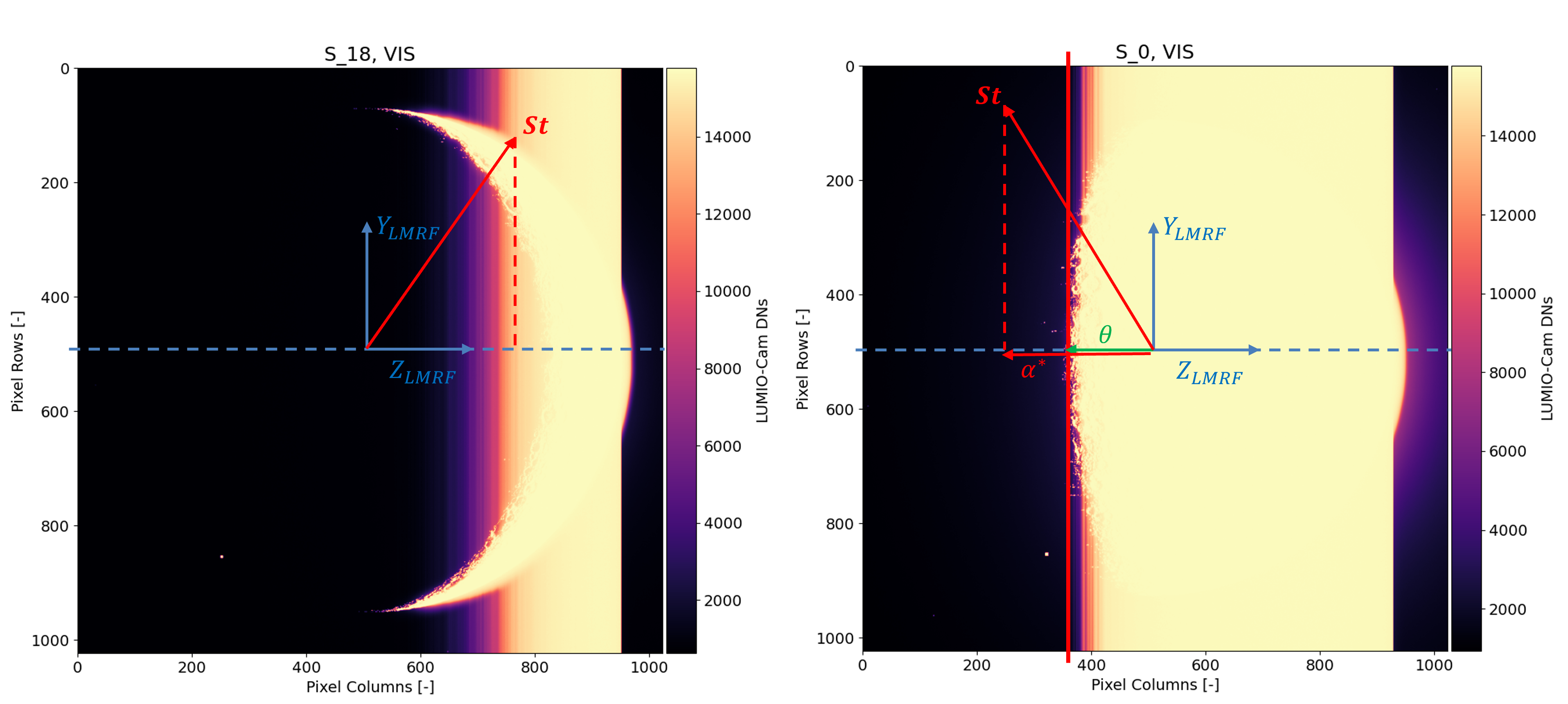}
    \caption{Occultation event selection based on the illuminating conditions. Left: high phase angle scenario; right: low phase angle scenario \cite{heywood-image}.}
    \label{fig:mesh3}
\end{figure}

Furthermore, the influence of the light reflected by the Moon's surface is modeled considering the position of the terminator plane with respect to the camera boresight. To this end, we defined the LUMIO-Moon Reference Frame (LMRF), which has its y-axis perpendicular to the plane containing the probe, the Moon, and the Sun, its x-axis pointing toward the Moon's center, and its positive z-axis in the Sun's direction.
As shown in Figure \ref{fig:mesh3}, two alternative scenarios might occur: when the Sun phase angle is below 90°, most of the Moon is illuminated; in this case, we need to define the angle between the x-axis and the terminator plane, $\theta$, and the angle between the x-axis and the projection of the star direction \textbf{St} on the x-z plane, $\alpha^*$. If the z-component of the \textbf{St} vector is positive, the occultation occurs on the illuminated side and the event is discarded. Similarly, when the z-component is negative but $\alpha^* < \theta$, occultation events are discarded due to the conservative assumption that the pixels representing this part of the Moon could be saturated, foreclosing the observation of the occultation. The second case occurs when the Sun phase angle is above 90° and most of the lunar surface is dark. For similar considerations, occultations occurring on the half perimeter of the Moon that is sunlit are discarded.
For the simulated time of occultations measurements, we assumed a conservative nominal uncertainty of 1 s that accounts for the LUMIO-Cam temporal resolution of 60 ms, and errors induced by the optics diffraction, extended light sources, and other unmodeled effects \cite{psiaki2007autonomous}.

Simulated observations include two-way range and Doppler radiometric measurements at X-band between LUMIO and the ESTRACK ground stations. Specifically, we assumed a 3-hour coverage at the beginning and end of each science cycle, corresponding to the scheduled correction maneuvers, and a 2-hour coverage in the middle. This schedule allows LUMIO to be tracked at least once per week, for a total tracking time of 8 hours every 15 days. For the range measurements, we assumed a conservative noise of 15 m. For the Doppler, we assumed a 0.1 mm/sec noise at 60 s count time, corresponding to a frequency variation of $2.8*10^{-3} Hz$.

\subsection{Dynamical Model}
All relevant forces acting on the spacecraft in the lunar environment were modeled within the OD setup. Specifically, for trajectory propagation, we considered: relativistic point-mass gravity from the Sun, the Solar System planets, and their moons; given the proximity to LUMIO, the Earth and Moon were modeled with spherical harmonics up to the degree and order 20, using respectively the EIGEN-GL04C and GL0660B models \cite{konopliv2013jpl,forste2008geoforschungszentrum}. The celestial bodies' mass parameters and state vectors were recovered from JPL’s DE440 planetary ephemerides \cite{park2021jpl}. The initial state of LUMIO was retrieved from the latest kernels, covering the period between September 2027 and February 2029, corresponding to the nominal lifetime of LUMIO plus a five-month extension.

\begin{table}[h]
    \centering
    \caption{Area and coefficients of LUMIO geometry.}
    \label{tab:my_label2}
    \begin{tabular}{|c|c|c|c|}
        \hline \textbf{Component} & \textbf{Area [$m^2$]} & \textbf{Reflectivity} & \textbf{Diffusivity}\\
        \hline
        Bus [front/top/side] & 0.04/0.06/0.06 & 0 & 0.25 \\
        \hline 
        Solar panels & 0.144 & 0.115 & 0.25 \\
        \hline
    \end{tabular}
\end{table}

We computed the non-gravitational acceleration due to solar radiation pressure (SRP) using a standard flat-plate model for a 12U CubeSat with deployed solar panels. Table \ref{tab:my_label2} summarizes the main assumed physical properties of the spacecraft surfaces. SRP models typically represent one of the primary error sources for the non-gravitational accelerations. We locally estimated SRP scale factors for each arc as part of the orbit determination process to mitigate possible modeling errors.
Furthermore, to simulate the unmodeled dynamics, we included stochastic accelerations acting in the spacecraft body axes using 8-hour interval batches and \textit{a priori} uncertainties of $10^{-12} km/s^2$ that are solved for within the filter (see Table \ref{tab:my_label3}).

\begin{table}[b!]
    \centering
    \caption{\textit{A priori} parameters and associated uncertainties used in the estimation filter.}
    \label{tab:my_label3}
    \renewcommand{\arraystretch}{1.3}
    \begin{tabularx}{\textwidth}{|p{2.2cm}|l|p{3cm}|X|}
        \hline
        \textbf{Parameter} & \textbf{Type} & \textbf{\textit{A priori} uncertainty ($\sigma$)} & \textbf{Comments} \\
        \hline
        
        \multicolumn{4}{|c|}{\textbf{LUMIO}} \\
        \hline
        Position & Local/Solve-for & $100\ km$ & Widely open \\
        \hline
        Velocity & Local/Solve-for & $0.1\ km/s$ & Widely open \\
        \hline
        Stochastic Accelerations & Local/Solve-for & $10^{-12}\ km/s^2$ & From personal communications with the navigation team \\
        \hline
        
        \multicolumn{4}{|c|}{\textbf{Moon}} \\
        \hline
        GM & Global/Consider & $1.4\cdot10^{-4}\ km^3/s^2$& \\
        \hline
        Radius & Local/Solve-for & $100\ m$& Considering a model like the one described by Fok et al. \cite{fok2011accuracy}. \\
        \hline
        
        \multicolumn{4}{|c|}{\textbf{Earth}} \\
        \hline
        GM & Global/Consider & $5\cdot10^{-4}\ km^3/s^2$& \\
        \hline
        
        \multicolumn{4}{|c|}{\textbf{SRP}} \\
        \hline
        Scale Factor & Local/Solve-for & $1.0$& The uncertainty is 100\% of the acceleration. \\
        \hline
        
        \multicolumn{4}{|c|}{\textbf{Star 141009 (example)}} \\
        \hline
        Position & Global/Consider & $8.54\cdot10^{-4}\ deg$& From UCACT-PI. Every uncertainty is referred to the right ascension and declination coordinates in the equatorial J2000 frame. \\
        \hline
        
        \multicolumn{4}{|c|}{\textbf{ESTRACK stations}} \\
        \hline
        Position & Global/Consider & $0.1\ m$& ESTRACK stations are Cebreros, Kourou, Malargue, New Norcia. \\
        \hline
        
        \multicolumn{4}{|c|}{\textbf{Media}} \\
        \hline
        Troposphere (wet/dry) & Global/Consider & $0.04\ m$ / $0.04\ m$ & The value is referred to all ESTRACK stations. \\
        \hline
        Ionosphere (day/night) & Global/Consider & $0.55\ m$ / $0.6\ m$ & The value is referred to all ESTRACK stations.\\
        \hline
        
        \multicolumn{4}{|c|}{\textbf{EOP (Earth Orientation Parameters)}} \\
        \hline
        Position & Global/Consider & $15\cdot10^{-9}\ rad$& \\
        \hline
        UT1 & Global/Consider & $25\cdot10^{-5}\ s$& \\
        \hline
    \end{tabularx}
\end{table}
\subsection{Filter Setup}
The parameter vectors of each arc include both solved-for and consider parameters. Solved-for parameters are estimated by the filter, while consider parameters are not directly estimated but are included to account for their contribution to the estimated uncertainty. In multi-arc estimation, parameters can also be classified into global parameters, common to all arcs, and local parameters, that vary and affect measurements within each arc \cite{godard2017multi}.
The estimation filter has been initialized with parameters resumed in Table \ref{tab:my_label3}.

\section{Results}
Two scenarios were considered for this analysis, which covers the LUMIO science cycles: the first one, indicated with \textit{Radiometric-only}, considers only radiometric observables obtained during tracking periods from ground stations. The second case, indicated with \textit{+Occultations}, considers radiometric observables and occultation events recorded throughout the entire arc. 

The LUMIO mission consists of about 22 science cycles. In the following section, only science cycles 20 and 18 are presented in detail, corresponding to the cases with the most and the least occultations per cycle duration. Overall statistics for the other science cycles will be presented as well. Results are shown in the Earth-centered Radial, Transverse, and Normal (RTN) reference frame to highlight the individual contributions of different observable types. The orbit determination performance in these two scenarios is evaluated by comparing their reconstructed position uncertainty. Similar considerations apply to the reconstructed velocity uncertainty, which is not reported here for brevity.

\subsection{Covariance analysis}

\begin{figure}[h]
    \centering
    \includegraphics[width=1\textwidth]{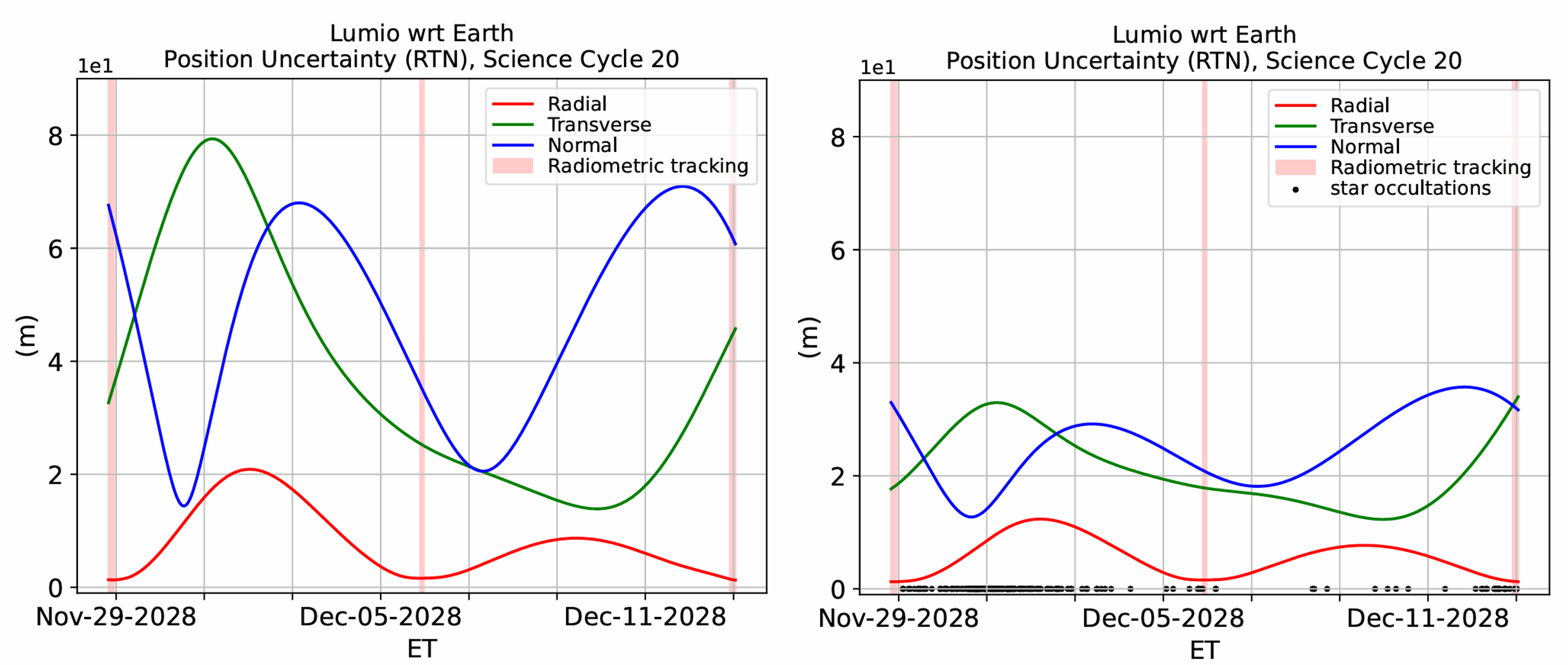}
    \caption{Estimated spacecraft position uncertainty (1$\sigma$) during science cycle 20. Left: \textit{Radiometric-only} scenario; right: \textit{+Occultations} scenario.}
    \label{fig:mesh4}
\end{figure}

Figure \ref{fig:mesh4} compares the estimated position uncertainty in the \textit{Radiometric-only} and \textit{+Occultations} scenarios for science cycle 20, where a total number of 197 occultation events are recorded. We can observe that the radial coordinate has a moderate improvement, with a 29\% average reduction, as summarized in Table \ref{tab:big_label}. Conversely, the transverse and normal coordinates improve by nearly a factor of two, with average uncertainty reductions of 43\% and 47\%, respectively. 
From Figure \ref{fig:mesh4}, we note that radiometric observables have an immediate effect on the radial component, where minima coincide with the periods of radiometric tracking. The effect on the normal coordinate seems delayed, with minima occurring later than the radiometric tracking periods. The transverse component seems mostly unaffected, with accuracies that improve at the epochs where star occultations are most concentrated.

\begin{figure}[htbp!]
    \centering
    \includegraphics[width=1\textwidth]{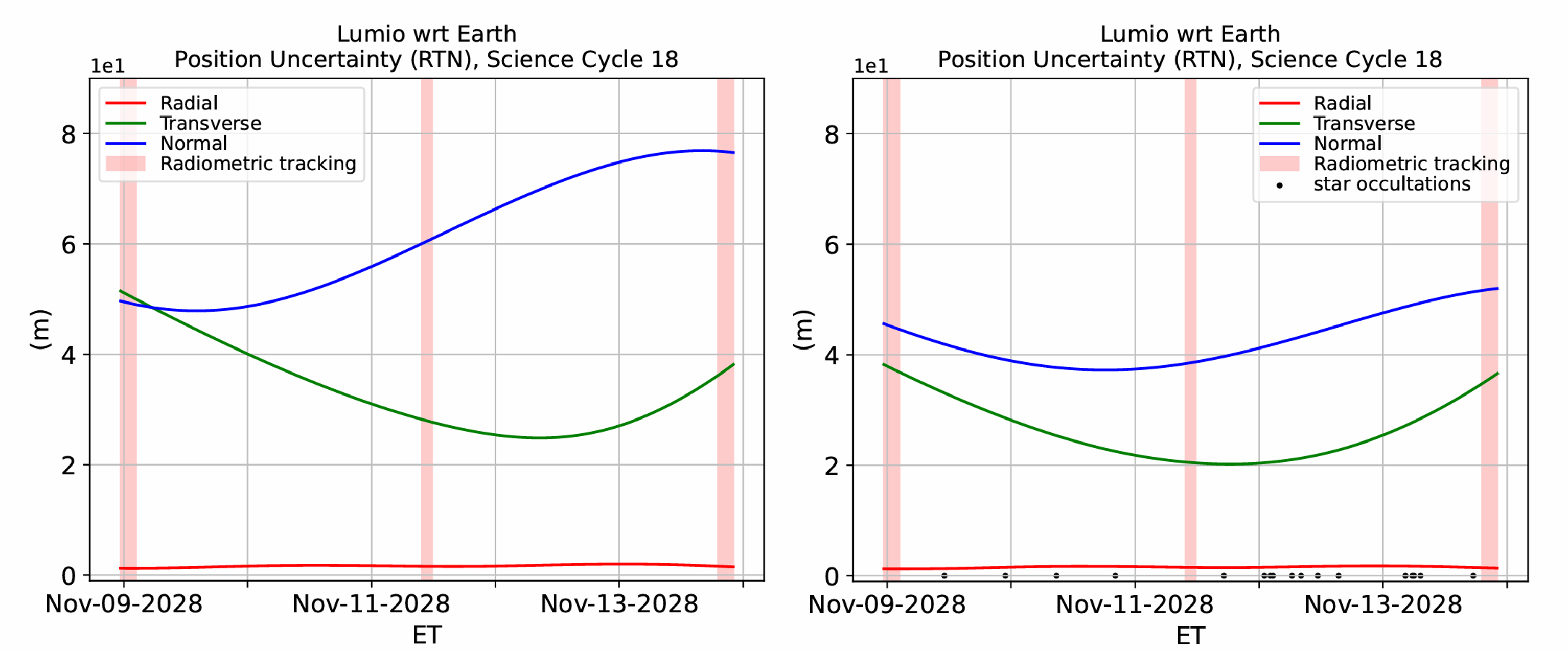}
    \caption{Estimated spacecraft position uncertainty (1$\sigma$) during science cycle 18. Left: \textit{Radiometric-only} scenario; right: \textit{+Occultations} scenario.}
    \label{fig:mesh5}
\end{figure}

Figure \ref{fig:mesh5} shows the estimated position uncertainties for science cycle 18, where 17 occultations events are recorded. In this case, for the \textit{+Occultations} scenario, we observe accuracy improvements in the order of 21\% and 31\% for the transverse and normal coordinates. Similarly to the previous case, the radial uncertainty is less affected, with an average relative improvement of 6\%, as shown in Figure \ref{fig:mesh5} and Table \ref{tab:big_label}.
The reduced relative improvement with respect to science cycle 20 highlights the dependence of the orbit determination performance on the total number of occultation measurements throughout the arc.

\begin{figure}[t!]
    \centering
    \includegraphics[width=1\textwidth]{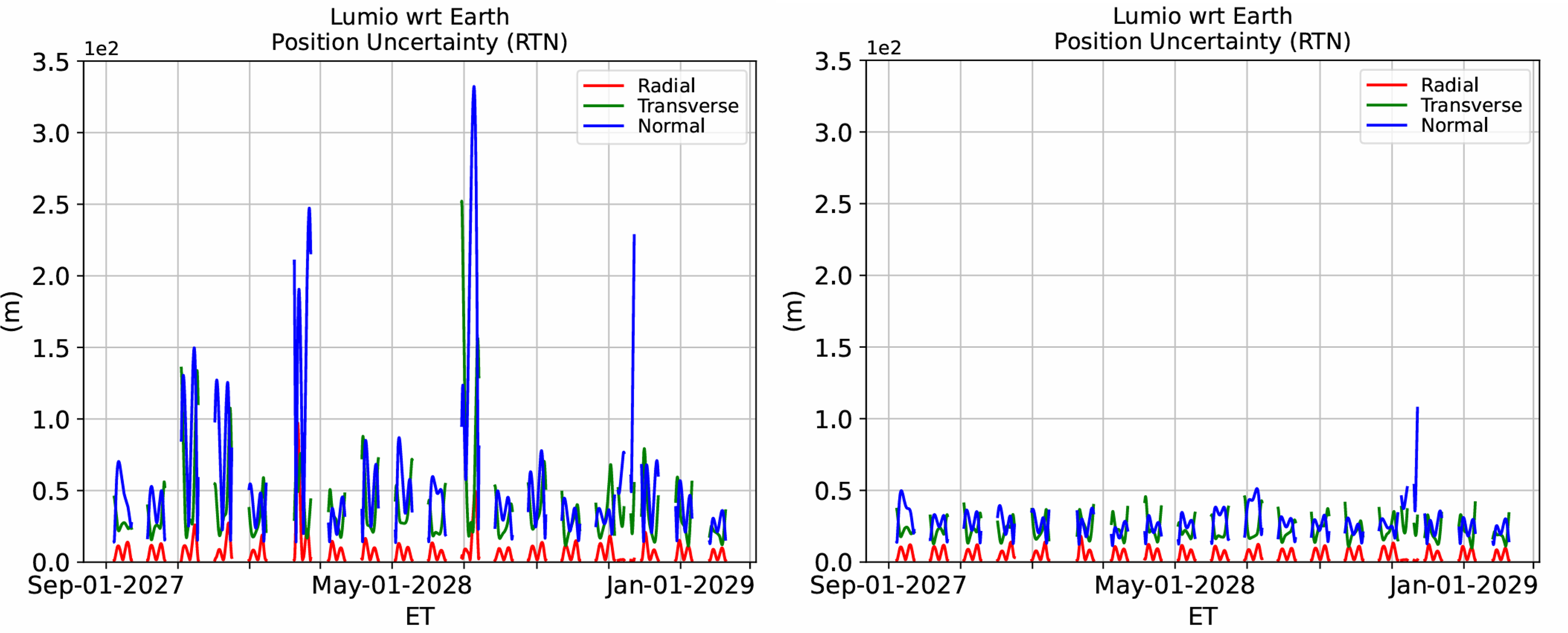}
    \caption{Estimated spacecraft position uncertainty (1$\sigma$) during all science cycles. Left: \textit{Radiometric-only} scenario; right: \textit{+Occultations} scenario.}
    \label{fig:mesh6}
\end{figure}

Figure \ref{fig:mesh6} shows the same comparison as the previous figures, but comprises all science cycles. There is a general improvement in the position uncertainty when occultation observables are introduced in the OD filter, particularly in the transverse and normal coordinates, which improve by up to one order of magnitude. Indeed, in Table \ref{tab:big_label}, the average transverse uncertainty decreases by 40\%, from 37 m to 22 m. The normal uncertainty decreases by 49\%, from 60 m to 30 m. Finally, the radial uncertainty decreases by 28\%, from 9.08 m to 6.55 m.

The higher return observed in the transverse and normal components is likely due to the nature of the range and Doppler radiometric observables, which provide observability mostly along the Earth-spacecraft line of sight. Precise time of occultation events place the spacecraft along a cylindrical surface whose axis is parallel to the star's direction and whose diameter makes the cylinder tangent to the Moon's surface, constraining the cross components of the spacecraft position in the Earth-centered RTN frame.

\begin{table}[t!]
    \centering
    \begin{threeparttable}
        \caption{Average position uncertainty of every science cycle during LUMIO's operational phase. Values are in [m].}
        \label{tab:big_label}
        \begin{tabular}{|c|c|c|c|c|c|c|c|c|c|}
            \hline
            \multicolumn{1}{|c|}{SC} & \multicolumn{3}{c|}{Radial} & \multicolumn{3}{c|}{Transverse} & \multicolumn{3}{c|}{Normal} \\
            \hline
            & Rad. & Rad.+Occ. & Ratio & Rad. & Rad.+Occ. & Ratio & Rad. & Rad.+Occ. & Ratio \\
            \hline
            01* & N/A & N/A & N/A & N/A & N/A & N/A & N/A & N/A & N/A \\
            \hline
            02 & 7.53 & 6.74 & 0.89 & 26.97 & 22.58 & 0.84 & 47.43 & 36.56 & 0.77 \\
            \hline
            03 & 7.24 & 6.66 & 0.92 & 28.44 & 21.78 & 0.77 & 38.96 & 27.06 & 0.69 \\
            \hline
            04 & 11.00 & 6.26 & 0.57 & 59.63 & 23.31 & 0.39 & 92.62 & 28.41 & 0.31 \\
            \hline
            05 & 10.16 & 6.47 & 0.64 & 45.64 & 22.76 & 0.50 & 83.23 & 27.75 & 0.33 \\
            \hline
            06 & 7.14 & 5.97 & 0.84 & 30.72 & 23.22 & 0.76 & 38.26 & 26.69 & 0.75 \\
            \hline
            07* & N/A & N/A & N/A & N/A & N/A & N/A & N/A & N/A & N/A \\
            \hline
            08 & 29.09 & 7.25 & 0.25 & 45.48 & 25.17 & 0.55 & 142.66 & 26.28 & 0.18 \\
            \hline
            09 & 7.14 & 5.94 & 0.83 & 31.72 & 23.39 & 0.74 & 31.33 & 21.50 & 0.69 \\
            \hline
            10 & 7.66 & 6.37 & 0.83 & 41.87 & 24.78 & 0.59 & 53.32 & 24.56 & 0.46 \\
            \hline
            11 & 6.87 & 5.97 & 0.88 & 38.78 & 21.75 & 0.56 & 54.24 & 26.39 & 0.49 \\
            \hline
            12 & 6.49 & 5.72 & 0.88 & 27.84 & 21.88 & 0.78 & 45.60 & 31.88 & 0.70 \\
            \hline
            13 & 17.38 & 6.36 & 0.36 & 86.07 & 24.52 & 0.28 & 180.75 & 40.07 & 0.22 \\
            \hline
            14 & 5.90 & 5.65 & 0.96 & 28.11 & 21.73 & 0.77 & 37.18 & 26.14 & 0.70 \\
            \hline
            15 & 7.30 & 6.26 & 0.86 & 39.60 & 24.94 & 0.63 & 49.64 & 25.58 & 0.51 \\
            \hline
            16 & 7.50 & 6.32 & 0.84 & 26.82 & 21.26 & 0.79 & 31.26 & 23.32 & 0.75 \\
            \hline
            17 & 8.60 & 6.92 & 0.80 & 35.02 & 23.24 & 0.66 & 31.57 & 25.12 & 0.80 \\
            \hline
            18 & 1.70 & 1.59 & 0.94 & 33.40 & 26.30 & 0.79 & 61.43 & 42.53 & 0.69 \\
            \hline
            19 & 1.31 & 1.17 & 0.89 & 30.30 & 22.55 & 0.74 & 121.62 & 59.90 & 0.49 \\
            \hline
            20 & 7.91 & 5.62 & 0.71 & 37.22 & 21.11 & 0.57 & 47.66 & 25.19 & 0.53 \\
            \hline
            21 & 7.67 & 6.15 & 0.80 & 32.39 & 22.18 & 0.68 & 37.55 & 24.00 & 0.64 \\
            \hline
            22 & 5.53 & 5.38 & 0.97 & 20.52 & 18.57 & 0.90 & 26.47 & 22.71 & 0.86 \\
            \hline
            \hline
            Avg. & 9.08 & 6.55 & 0.72 & 37.78 & 22.55 & 0.60 & 60.00 & 30.70 & 0.51 \\
            \hline
        \end{tabular}
        \begin{tablenotes}
            \small
            \item Note: The "Ratio" column indicates the second value divided by the first for each coordinate.
            \item * This arc had no visible occulted stars at the assumed magnitude.
        \end{tablenotes}
    \end{threeparttable}
\end{table}

\subsection{Sensitivity analyses}
In this paragraph, we assess how robust the technique is to errors in occultation time measurements and lunar shape model, with the intent to extend the technique to operational scenarios other than LUMIO, which might have different sampling times for the camera, noise sources, or different \textit{a priori} knowledge of the target shape. 
Science cycle 20 will be analyzed in detail, while the performance related to other science cycles will be summarized through average statistics. Being science cycle 20 the one with the most concentrated occultation events, it is in fact the most suited for highlighting performance variations due to changes in the values of relevant parameters. 

\subsubsection{Occultation measurements noise\\}
The \textit{+Occultations} scenario has been tested by varying the noise values of the occultation events between 100 ms, 1 s, and 10 s, keeping all other parameters fixed at the values described before. Variations in measurement noise are associated with those error sources that are not directly accounted for in the OD filter, such as: time resolution of the lightcurve extraction (i.e., camera sampling rate), jitter of the onboard clock, edge diffraction of the starlight as it passes the Lunar limb. Other error sources, such as long-term linear drift of the onboard clock, Lunar limb position uncertainty, and reported star coordinates in the plane of the sky, are already accounted for in the OD filter, either as solved-for or consider parameters.
The value of 1 s adopted in the previous sections is referred to as the \textit{nominal} scenario. 

\begin{figure}[t!]
    \centering
    \includegraphics[width=1\textwidth]{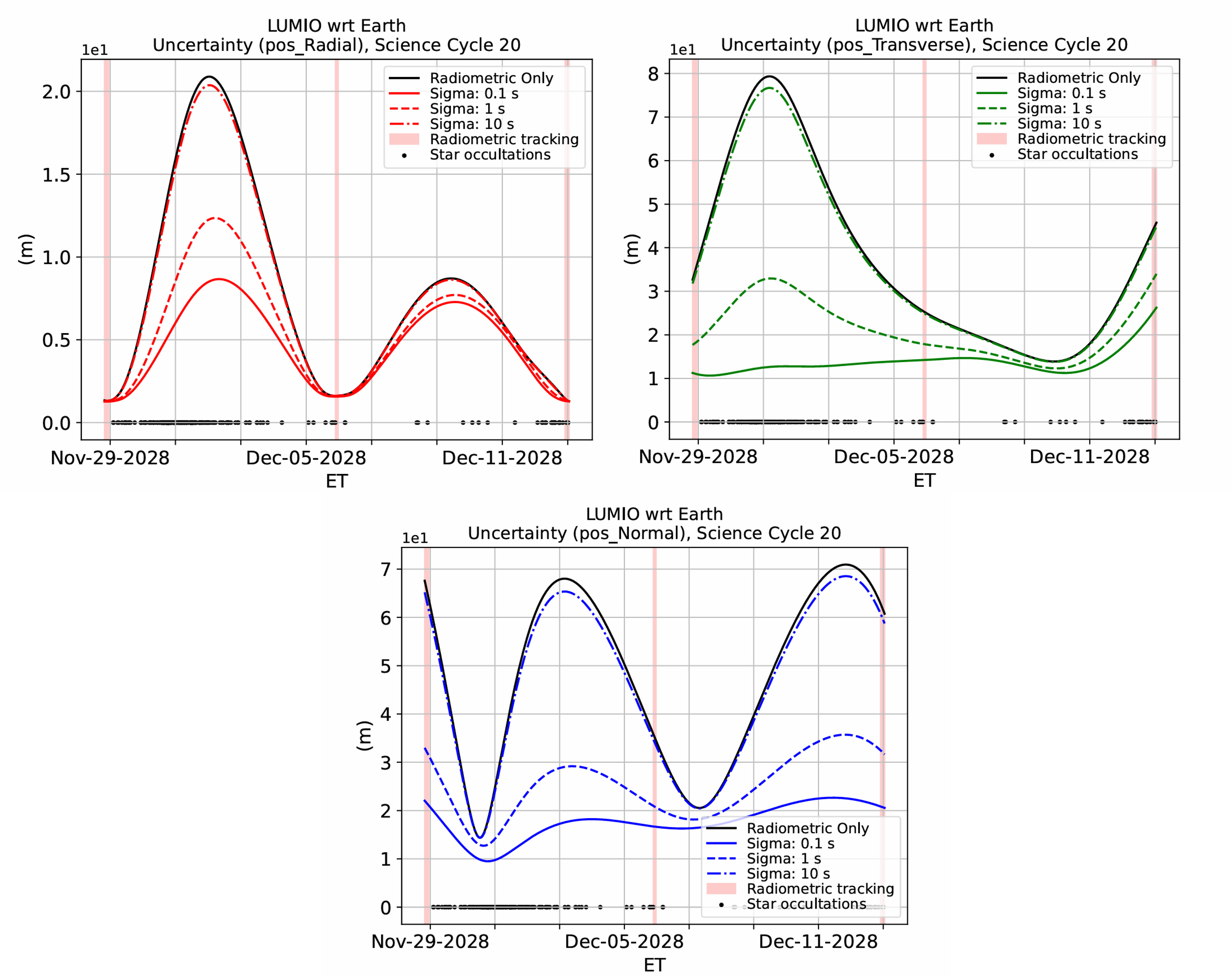}
    \caption{Estimated spacecraft position uncertainty \(1\sigma\) for science cycle 20 in the radial (red), transverse (green), and normal (blue) directions as a function of the occultation measurement noise in the \textit{+Occultations} case: 10 s (dash-dotted line), 1 s (dashed line), and 100 ms (continuous line). The black line reports the \textit{Radiometrics-only} case for reference.}
    \label{fig:mesh7}
\end{figure}

Figure \ref{fig:mesh7} shows the estimated position uncertainty as a function of time and measurement noise values through science cycle 20. The formal uncertainty obtained for the \textit{Radiometrics-only} scenario is also provided as reference. For the \textit{+Occultations} scenario, the radial component shows maximum relative improvements ranging between 2\% (\({\sigma}_{occ}=10\ s\)) and 40\% (\({\sigma}_{occ}=0.1\ s\)), which occur during periods when no radiometric tracking is performed and the filter relies solely on the knowledge of the spacecraft dynamics for state propagation.
The transverse component shows a more consistent improvement throughout the arc, with maximum relative improvements ranging between 2\% and 63\%, respectively. For measurement noise of 10 s, the observability of this coordinate is mostly defined by the spacecraft dynamics and radiometric data. As the measurement error is reduced, the difference with the \textit{Radiometrics-only} curve is larger, and accuracy improvements are concentrated in periods of frequent star occultations. Similar considerations apply for the normal component, whose maximum relative improvement ranges between 3\% and 63\%.
Average values of the position uncertainty over the whole arc are also reported in Table \ref{tab:my_label4}. 

Even though the accuracy improvement is smaller for timing errors $>$1 s, we can still observe an appreciable contribution. This could indicate the potential to extend this technique to other operational conditions that foresee longer camera exposure times. For example, commercial off-the-shelf star trackers typically operate with sampling rates between 1 Hz to 10 Hz \cite{zhu_ch8_2022}.

\begin{table}[t!]
    \centering
    \caption{Average LUMIO position uncertainty as a function of the occultation measurement noise for science cycle 20. Values are provided in the Earth-centered RTN frame.}
    \label{tab:my_label4}
    \begin{tabular}{|l|c|c|c|}
        \hline
        \textbf{} & \textbf{Radial [m]} & \textbf{Transverse [m]} & \textbf{Normal [m]} \\
        \hline
        \textbf{Radiometrics}      & 7.91 & 37.22 & 47.67 \\
        \hline
        \textbf{Sigma: 10 s}       & 7.76 & 36.31 & 46.12 \\
        \hline
        \textbf{Sigma: 1 s}        & 5.62 & 21.11 & 25.19 \\
        \hline
        \textbf{Sigma: 100 ms}     & 4.68 & 13.67 & 17.57 \\
        \hline
    \end{tabular}
\end{table}

\subsubsection{Moon shape uncertainty\\}
A second sensitivity test has been made by varying the \textit{a priori} uncertainty on the Moon shape, while keeping the occultation measurement noise fixed at 1 s. The tested values are 10 m, 100 m, and 1 km. The first two values represent the best and worst expected uncertainty from the currently available digital maps of the Moon retrieved by the Lunar Reconnaissance Orbiter (LRO) mission \cite{fok2011accuracy, barker2016new}. The 100 m uncertainty value, which is also used in the nominal covariance analysis, represents a conservative assumption for assessing the OD performance in a post-flight scenario, where high-accuracy digital elevation maps can be processed on the ground. The 1 km uncertainty case represents a highly conservative scenario that is more consistent with the use of lower-fidelity Moon shape models for real-time processing onboard the spacecraft for autonomous applications.

\begin{figure}[t!]
    \centering 
    \includegraphics[width=1\textwidth]{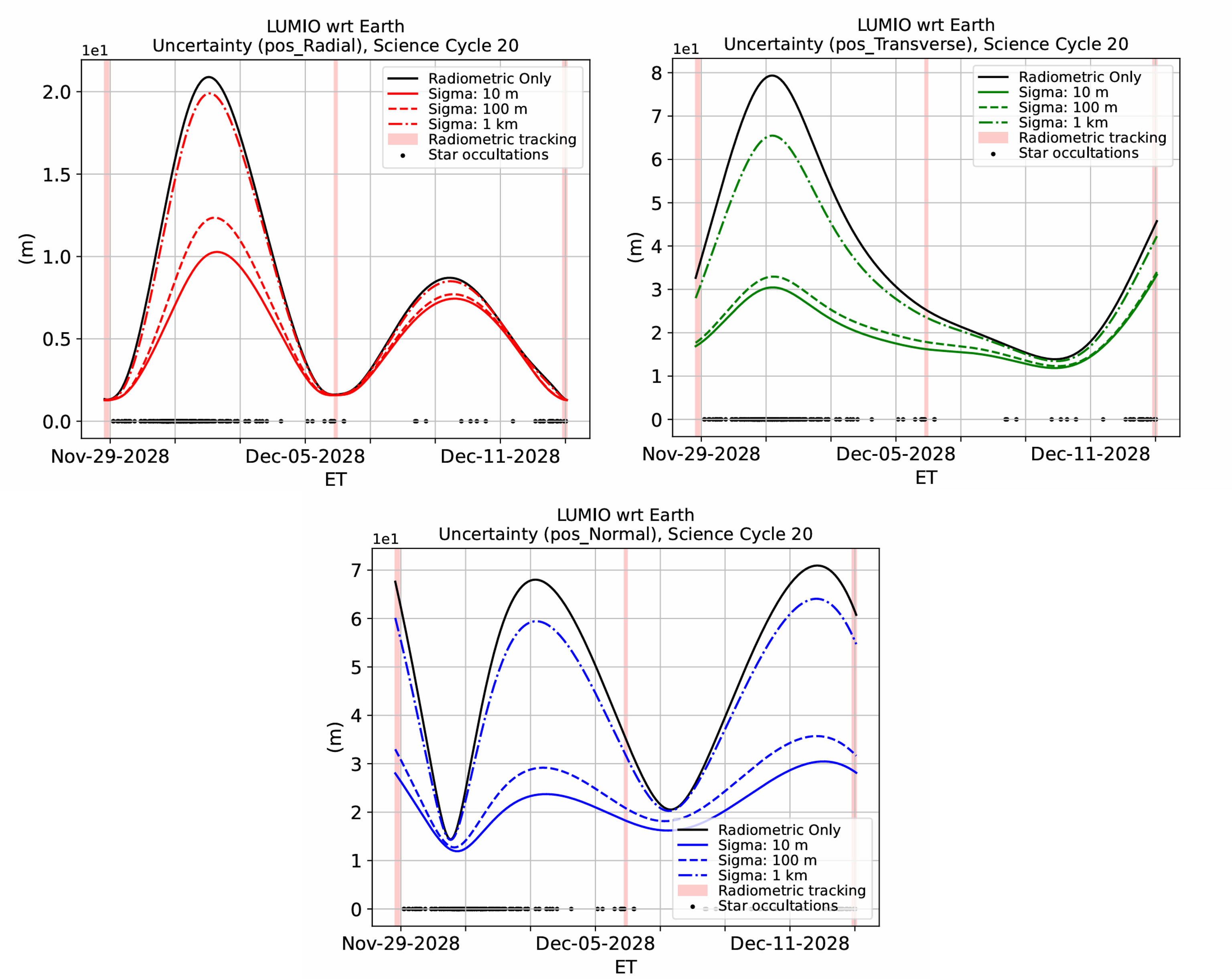}
    \caption{Estimated spacecraft position uncertainty \(1\sigma\) for science cycle 20 in the radial (red), transverse (green), and normal (blue) directions as a function of the Moon shape uncertainty in the \textit{+Occultations} case: 1 km (dash-dotted line), 100 m (dashed line), and 10 m (continuous line). The black line reports the \textit{Radiometrics-only} case for reference. }
    \label{fig:mesh8}
\end{figure}

\begin{table}[t!]
    \centering
    \caption{Average LUMIO position uncertainty as a function of the Moon's shape uncertainty for science cycle 20. Values are provided in the Earth-centered RTN frame.}
    \label{tab:my_label5}
    \begin{tabular}{|l|c|c|c|}
        \hline
        \textbf{} & \textbf{Radial [m]} & \textbf{Transverse [m]} & \textbf{Normal [m]} \\
        \hline
        \textbf{Radiometrics}      & 7.91 & 37.22 & 47.67 \\
        \hline
        \textbf{Sigma: 1 km}       & 7.57 & 32.56 & 42.99 \\
        \hline
        \textbf{Sigma: 100 m}      & 5.62 & 21.11 & 25.19 \\
        \hline
        \textbf{Sigma: 10 m}       & 5.11 & 19.75 & 21.41 \\
        \hline
    \end{tabular}
\end{table}

Figure \ref{fig:mesh8} shows the estimated position uncertainty as a function of time and lunar shape uncertainty for science cycle 20. The formal uncertainty obtained for the \textit{Radiometrics-only} scenario is also provided as reference. Similarly to the previous analysis, the radial uncertainty lowers from 2\% (\({\sigma}_{shape} = 10\ m\)) to 35\% (\({\sigma}_{shape} = 1\ km\)), deviating from the \textit{Radiometrics-only} case mostly during periods of no tracking, where the accuracy is degraded. 
The transverse component has a maximum relative improvement ranging between 13\% and 47\%, while the normal component shows maximum relative improvement between 10\% and 55\%. Average uncertainty values over the whole arc are also reported in Table \ref{tab:my_label5}.
Although the curves corresponding to different shape uncertainties are closer to one another, the time dependence is similar to that observed for the measurement error sensitivity. This stems from the fact that shape errors translate directly into occultation time errors.

The analyzed shape uncertainties are consistent with absolute uncertainties of real small bodies. For example the estimated uncertainty of Didymos' equivalent diameter before DART's impact was of 17 m  \cite{richardson2024dynamical}. Similarly, Apophis (99942) shape has been derived through radar observations and the uncertainty of its equivalent diameter is roughly 40 m  \cite{brozovic2018goldstone}. This observation hints at the possibility of extending the current concept to deep space missions towards small bodies, even during early mission phases, when coarse shape models are typically available.

\section{Conclusions}
This work presented a feasibility analysis on the use of stellar occultations to improve orbit determination accuracy for the Lunar Meteoroid Impact Observer (LUMIO), a CubeSat operating in a quasi-Halo orbit around the Earth–Moon L2 point. The focus was on performance during LUMIO’s science phases, when radiometric tracking is limited and low illumination prevents traditional optical navigation techniques from being used.
Stellar occultation events, defined as precise timings of stars passing behind the Moon’s limb, were simulated using a Python-based tool leveraging JPL’s MONTE library. Non-viable events were excluded based on geometric and illumination constraints, including field-of-view limitations and glare from the Sun, the Earth, and the illuminated lunar surface. These observables were incorporated into a batch least-squares filter to estimate the spacecraft’s state.
Covariance analysis showed that augmenting radiometric data with occultation measurements improves the position estimation, particularly in the transverse and normal components, with uncertainty reductions up to a factor of two. These improvements are most significant during arcs with a high density of occultations and during intervals without radiometric tracking, where the filter depends more heavily on dynamics.
Sensitivity studies revealed that the OD performance is strongly influenced by the timing precision of occultation measurements, with sub-second accuracy enabling substantial gains. Conversely, the impact of lunar shape uncertainty is minimal below 100 m, supporting practical application even with moderately accurate topographic models. This has positive implications for extending the approach to autonomous onboard navigation and missions targeting small bodies, where the shape knowledge is limited.
Although the current analysis uses a simplified occultation model, its conservative assumptions make the results representative of realistic mission conditions. Future work should incorporate high-fidelity lunar topography, optimize the star catalog selection based on exposure times, and refine the error budget to include effects such as clock jitter and limb diffraction. Given the enormous volume of data expected from the LUMIO-Cam during science observations, onboard processing and data compression will also be critical for operational viability.
Finally, future work shall address the implications of the improved OD performance in terms of surface localization of the observed LIF events, which represent the main scientific goal of the mission and could benefit from improved spacecraft constraints.

\section{Acknowledgment}
DB, RLM, and MZ acknowledge Caltech and the NASA Jet Propulsion Laboratory for granting the University of Bologna a license to an executable version of MONTE Project Edition S/W. DB, RLM, and MZ are grateful to the Italian Space Agency (ASI) for financial support through Agreement No. 2024-6-HH.0 in the context of ESA's LUMIO mission. The authors also acknowledge Francesco Topputo and Fabio Ferrari from Politecnico di Milano for their outstanding role in coordinating the LUMIO mission and the science working groups.

\bibliographystyle{AAS_publication}   
\bibliography{references}   

\end{document}